\documentclass[aps,showpacs,a4paper,floatfix,twocolumn,prc,amsmath,amssymb]{revtex4-1}
\usepackage{amsmath}
\usepackage{graphicx}
\usepackage{ulem}
\usepackage{morefloats}
\usepackage{rotating}
\usepackage{relsize}
\usepackage{color}
\usepackage{floatflt,epsfig}
\usepackage{ulem}
\usepackage{natbib}
\usepackage{hyperref}
\usepackage{filecontents}

\begin{document}

\title{Can generalized TOV equations explain neutron star macroscopic properties?}

\author{ Cl\'esio E. Mota $^2$, Guilherme Grams$^2$ and D\'ebora P. Menezes $^2$}
\affiliation{
$^2$ Departamento de Física, Universidade Federal de Santa Catarina,
Florianópolis, SC, CP 476, CEP 88.040-900, 
Brazil\\
}

\begin{abstract}
In the quest for describing massive neutron stars with radii compatible with recent expectations, we take a different approach in the present paper and, instead of tuning the equations of state, we use an appropriate parameterization of the TOV equations that result in increasing the maximum mass and decreasing the respective radius and consequently, also the radius of canonical stars.
\end{abstract}

\maketitle

\section{Introduction}

The understanding of the complete quantum chromodynamics (QCD) phase
diagram is not a trivial task and depends on the development of
physics based on different strategies according to the region of
interest. These strategies vary from lattice QCD, heavy ion
collision experiments to the study of compact objects, the latter
being  the motivation for the present work.   

The physics of compact objects involves three basic steps, namely: i)
the construction of an appropriate equation of state (EoS), ii) the
use of the Tolman-Oppenheimer-Volkof equations (TOV), which
are relativistic hydrostatic equilibrium equations \cite{tov1,tov2} and iii) 
the comparison of the macroscopic properties with the ones observed by 
astrophysicists.

The EoS can be obtained from different formalisms, but the two most
common groups are the Skyrme-like models and hence, non
relativistic and the group originated from mean-field theory applied to
relativistic lagrangian densities. Both non-relativistic and
relativistic models are written in terms of parameters that are fitted
to reproduce bulk nuclear matter and some times, also finite nuclei
properties. In the last 5 or 6 decades, hundreds of models were
proposed and largely utilized. Important bulk properties are the
binding energy, incompressibility, symmetry energy
and its slope, all of them calculated at the nuclear saturation density.
In general, values for the symmetry energy and
its slope come from heavy ion collision experiments, pygmy dipole resonances,
isobaric analog states and isoscalar giant monopole resonances.
Incompressibility values are obtained from isoscalar giant monopole
resonances and isovector giant dipole resonances and its volume part from neutron skin thickness and
isospin diffusion calculations.

A common problem of Skyrme models are the fact that
some EoS become acausal and the symmetry energy decrease too  much beyond
three times saturation density, which is a serious deficiency if one
wants to apply the EoS to describe neutron stars. In 2012, 240 different Skyrme model
parameterizations were confronted with experimentally and empirically
derived constraints. Only 16 models were considered approved
\cite{Dutra2012}. 

Soon afterwards, 263 relativistic mean field models (RMF) models were also
confronted with the same {\it adapted} experimental constraints and
once again the vast majority was shown not to satisfy all necessary conditions
to be approved. Only 35 of them satisfied all constraints \cite{Dutra2014}.
 
From the astrophysical side, until 2010, all EoS used as input to the 
TOV equations that resulted in a family of stars with a maximum mass
star larger than the canonical neutron star, which has a 1.4 $M_\odot$,
were accepted. However in 2010 and 2013 two massive stars were
discovered, namely, PSR J1614-2230 with mass ($1.97 \pm 0.04$) $M_\odot$
\cite{Demorest}, and PSR J0348+0432 with mass ($2.01 \pm 0.04$) $M_\odot$
\cite{Antoniadis} and a very stringent constraint was born. 
It is worth pointing out that models designed to be applied
around nuclear saturation density are not expected to remain valid at 
densities typical of neutron star interiors, which can be up to 6 times higher. 
Hence, the validity of both Skyrme-type and RMF models has to be checked 
on an individual basis.
Since the detection of the above mentioned massive stars, 
different hypotheses lead to predictions of the radii
of the canonical neutron star varying from 9.7-13.9 km
\cite{Hebeler2013}  to 10.4-12.9 km \cite{Steiner2013} 
and from 10.1 to 11.1 km \cite{Ozel2016}. Some of these results point
to vary small radii, since in general, the canonical star has a radius
larger than the maximum mass star. Hence, another restrictive
constraint became important, despite the fact that the calculations 
that lead to these radii bounds carry big systematic uncertainties,
not always mentioned.

Besides the individual Skyrme type parameterizations, 
a meta-model with underlying non-relativistic model
assumptions  was developed \cite{francesca} and 25 million EoS
were tested against direct URCA processes. Approximately 4 million satisfied all
physical requirements. These EoS were then analyzed with respect to
the neutron star radii.  The authors concluded that neutron stars
constituted exclusively of nucleons and leptons with masses between 1
and 2 $M_\odot$ have radii of $12.7 \pm 0.4$ km, i.e., incompatible
with most of the predictions based on astrophysical analyses mentioned
above. It is fair to mention that these narrow interval for the radii
 results from the choice the authors made for the priors. The radii
 bounds could be larger, had a different choice being made. 

Moreover, the same 35 models shown to successfully describe nuclear matter
properties in \cite{Dutra2014} were investigated with respect to
stellar matter properties in \cite{Dutra2016}.  Only 12
parameterizations resulted in neutron stars with maximum mass in the
range of 1.93-2.05 $M_\odot$ and the conclusions about the radii size
corroborated the ones found in \cite{francesca}. 
If hyperons are included in the calculations, the situation becomes even more
complicated because the EoS must be soft at subsaturation densities
and hard at higher densities to predict massive stars, but hyperons
soften the EoS. In this case, to reconcile the recent measurements of massive stars
with relatively small radii, either strange mesons or a new degree of freedom
(not necessarily  known) has to be incorporated in the calculations
\cite{Luiz2018}. Moreover, all models that include hyperons
depend on hyperon-meson couplings, largely unknown. Generally, the
prescriptions used are based on symmetries (SU(3) or SU(6)) or on
the fitting of phenomenological potentials. However, in a dense matter
as the one existing in the interior of neutron stars, hyperons are
indeed expected to appear. We come back to this point when we
discuss the formalism adopted in the present work.

More recently, also the 16 Skyrme-type models approved in 
\cite{Dutra2012} were confronted with astrophysical constraints 
\cite{cskyrme} and 5 of them were shown to describe massive stars 
and bear radii inside the bands obtained in \cite{Lattimer2016}, 
which resulted from a Bayesian analysis of type-I X-ray burst 
observations. 

In 2017, the gravitational wave  GW170817 was detected as the
result of the merging of two neutron stars belonging to a binary
system \cite{GW170817}. The tidal deformability data, which provided 
a relation between the neutron star inner structure and the emitted 
gravitational wave, offered a new opportunity to constrain equations of state. 
Since then, many EOS were used to compute tidal deformability parameters
for the case of the low and the high mass NS components of the binary
system. In \cite{cskyrme}, the very same 16 Skyrme-type models approved in
\cite{Dutra2012} were tested and 4 out of the 5 models that could also
describe massive stars, were inside the accepted confidence lines
interval obtained from the tidal polarizabilities of the two NS binary system components. 
The radii of canonical stars described by these 4 consistent 
Skyrme parameterizations lie in the interval $11.82 < R_{1.4} < 12.11$ km, 
values which are compatible with the ones found in \cite{malik}. 
The very same approach was adopted \cite{crmf} with respect to 34 RMF models approved in 
\cite{Dutra2014}. Although all 34 models lie in between the confidence lines mentioned above,
and 24 of them give canonical star radii compatible with the predictions given in \cite{Ligo2},
only 5 consistent relativistic mean field models were found to simultaneously 
describe massive stars and constraints from GW170817.

As a summary, systematic studies of more than 500 models have shown that only 4 out of 16 consistent Skyrme models and only 5 out of 34 consistent RMF models (none of them including hyperons) are reliable to 
describe massive stars and recently accepted astrophysical constraints. We then wonder...what if the problem lies in the use of the TOV equations and
not in the choice of the EoS? 

A possible way of circumventing this paradox was proposed not long ago and it was based on the solution of TOV-like equations in the braneworld \cite{benito2014}. In that work, it was shown that it is possible to reconcile large masses and small radii with ordinary equations of state, without the need of extra flamboyant degrees of freedom. However, the price paid was the need to solve 4 coupled equations of state with two extra parameters and to rely on the existence of the exotic braneworld. In the present paper we try a more standard approach, i.e., the original TOV equations are solved, but some free parameters are included as part of their expressions, all of them coming from the relativistic hydrostatic equilibrium equations.

\section{Formalism}

\subsection{TOLMAN - OPPENHEIMER - VOLKOFF EQUATIONS - TOV}
In this section we review some of the main procedures that leads to the 
deduction of the TOV equations.

Einstein's equations are the basis for the theory of gravitation, known as General Relativity (GR).
Through a suitable treatment, the solutions of these equations allow us to obtain a description of the
hydrostatic equilibrium of homogeneous, static (no rotation), isotropic and spherically symmetrical objects \cite{Carroll2004}, such as neutron stars, white dwarfs, magnetars and others. Also known as compact objects, in astrophysics. In the description of these objects, the Einstein equations are treated in static and isotropic regions of space-time found in the outer and inner regions of compact stars treated roughly as perfect fluids. Only the stellar interior is treated in the present discussion.

Based on the symmetries mentioned above, we must define a metric in an appropriate coordinate system that describes the object being studied. The most general metric describing space-time under consideration is given by the line element

\begin{equation}
ds^{2}=B(r)dt^2-A(r)dr^2-r^2(d\theta^2+\sin^2\theta d\phi^2),
\label{16}
\end{equation}
where we define $B \equiv B(r)$ and $A \equiv A(r)$, are functions that we want to determine based on the field equations. We are assuming the metric signature $(+---)$ and the unit convention $c = 1$. 

Therefore, the metric covariant tensor $g_{\mu\nu}$ has the following non-null components: 

\begin{align}
g_{00}&= B(r) \ , \ \  g_{11}=-A(r) \ , \ \ g_{22}=-r^2 \ , \nonumber \\ 
g_{33}&=-r^2 \sin^2 \theta\ , 
\end{align}
together with its contravariant version $g^{\mu\nu}$:
\begin{align}
g^{00}&= \frac{1}{B(r)} \ , \ \ g^{11}=\frac{-1}{A(r)} \ , \ \ g^{22} =\frac{-1}{r^2} \ , \nonumber \\  g^{33}&=\frac{-1}{r^2 \sin^2 \theta}.
\end{align}

There are other ways of writing the equation (\ref{16}), for example, by defining $B(r)=e^{2\Psi(r)}$ and $A(r)=e^{2\Phi(r)}$. We keep the general form, $B(r)$ and $A(r)$, in the course of our derivations to facilitate comparison with the Newtonian potential. 

The Einstein field equations for the stellar interior is given by:
\begin{equation}
G_{\mu\nu}=R_{\mu\nu}-\frac{1}{2}g_{\mu\nu}R = k T_{\mu\nu}.
\label{2}
\end{equation}
where $k = 8\pi G$ is the Einstein gravitational constant, chosen to maintain agreement
with the Newtonian limit. $ R_{\mu\nu}$ is the Ricci tensor, $R$ is the scalar of curvature 
(trace of $R_{\mu\nu}$), $ G_{\mu\nu}$ is the Einstein geometric tensor and $ T_{\mu\nu}$ is the energy-momentum tensor responsible for describing matter in the stellar interior. Taking into account that we can model the matter of the star as a perfect fluid, the energy-momentum tensor in a comoving frame can be written as
\begin{equation}
T_{\mu\nu}=(\rho+p)u_{\mu}u_{\nu}-pg_{\mu\nu},
\label{3}
\end{equation}
where respectively $\rho $ and $p$ represents the density and pressure of the fluid and are functions only of the radial coordinate $r$. The connection between pressure and density is provided by means of an appropriate equation of state. The term $u_{\mu}\equiv dx_{\mu}/d\tau$ is defined as the 4-velocity of a fluid element, satisfying a normalization condition $(u_{\mu}u^{\mu}=-1)$ and, in addition
\begin{equation}
u_{\mu}=(B,0,0,0).
\end{equation}
Therefore, the $T_{\mu\nu}$ components, for the $u_{\mu}$, are given by
\begin{align}
T_{00} &= \rho B(r) \ , \  \ T_{11}=p A(r) \ , \ \ T_{22} = p r^2 \ , \nonumber \\
T_{33} &= p r^2 \sin^2 \theta \ , 
\end{align}
where in turn the contributions
\begin{align}
T^{00}&= \frac{1}{\rho B(r)} \ , \ \ T^{11}=\frac{1}{p A(r)} \ , \ \ T^{22} =\frac{1}{p r^2} \ , \nonumber \\
T^{33}&=\frac{1}{p r^2 \sin^2 \theta} \ ,
\end{align} 
are the components of $T^{\mu\nu}$. Based on the results described above, we can write the components of $G_{\mu\nu}$. Therefore, we have in effect the expressions:

\begin{equation}
G_{00}= \frac{A'B}{rA^{2}}+\frac{B}{r^{2}}\left( 1-\frac{1}{A}\right), 
\end{equation}

\begin{equation}
G_{11}= \frac{B'}{rB}-\frac{A}{r^{2}}\left( 1-\frac{1}{A}\right), 
\end{equation}

{\fontsize{9.5}{12}
\begin{align}
G_{22}&= \frac{r^{2}B''}{2AB}-\frac{r^{2}B'}{4AB}\left( \frac{A'}{A}+\frac{B'}{B}\right) \nonumber \\
  &-\frac{r}{2A} \left(\frac{A'}{A}-\frac{B'}{B}\right), 
\end{align}
}

\begin{equation}
G_{33}= G_{22} \sin^{2} \theta, 
\end{equation}
where, the prime denotes a derivative with respect to the radial coordinate $r$. We can observe that $G_{\mu\nu}$ is diagonal, which indicates characteristics of an isotropic space, which in turn implies homogeneity, being in agreement with the space-time symmetries initially imposed. Now we can write the Einstein equations (\ref{2}), to describe how a spherically symmetrical distribution of matter considered as a perfect fluid deforms the space in its interior. Thus, by combining the results of the components for $G_{\mu\nu}$ and for the energy-momentum tensor $T_{\mu\nu}$, we obtain 3 coupled differential contributions:

\begin{equation}
\frac{A'B}{rA^{2}}+\frac{B}{r^{2}}\left( 1-\frac{1}{A}\right) = 8 \pi G B \rho, 
\label{17} 
\end{equation}

\begin{equation}
\frac{B'}{rB}-\frac{A}{r^{2}}\left( 1-\frac{1}{A}\right)=8 \pi G A p, 
\label{18}
\end{equation}

{\fontsize{9.5}{12}
\begin{align}
\frac{r^{2}B''}{2AB}-\frac{r^{2}B'}{4AB}\left( \frac{A'}{A}+\frac{B'}{B}\right)&-\frac{r}{2A}\left(\frac{A'}{A}-\frac{B'}{B}\right) \nonumber \label{19} \\
&=8 \pi G r^{2}p.
\end{align}}
\noindent The integration of equation (\ref{17}) leads to
\begin{equation}
A(r)=\left( 1-\frac{2GM(r)}{r}\right)^{-1},
\label{20}
\end{equation}
together with the definition
\begin{equation}
\small M(r)\equiv \int_{0}^{r} 4 \pi r^2 \rho(r)dr.
\label{1}
\end{equation}
according to the above definition for $ M(r)$, the integration of energy density $ \rho(r) $ occurs in the stellar interior, and can thus be interpreted as the mass contained within $r$. The definition (\ref{1}) is known as the continuity equation for mass and expressed how the mass inside a star varies according to the radius. It is formally the same obtained from a non-relativistic treatment. We consider that $ r = R $ denotes the radius of the star, that is, a radial coordinate at which the pressure disappears,
\begin{equation}
\small M_{G}\equiv M(R)= \int_{0}^{R} 4 \pi r^2 \rho(r) dr,
\label{4}
\end{equation}
is the gravitational mass of the object measured by a distant observer from gravitational effects. It should be emphasized that throughout this work $M(R)$ is called simply mass, indicated by the letter $M$.

Manipulating the equation (\ref{18}), employing the result (\ref{20}) together with the definition $B(r)=e^{2\Psi(r)}$, we can write an expression for the gravitational potential:

{\fontsize{9.4}{12}
\begin{align}
\frac{B'(r)}{2B(r)} \equiv \Psi'(r) = \frac{GM(r)}{r^{2}}\left(1+\frac{4\pi r^{3} p(r)}{M(r)}\right) \left( 1 - \frac{2GM(r)}{r}\right)^{-1}. \label{21}
\end{align}
}
At this stage making use of the conservation law for the energy-momentum tensor $\nabla_{\mu}T^{\mu\nu} = 0$ (where $\nabla_{\mu}$ denotes the covariant derivative) and obtain the relation
\begin{equation}
\Psi'(r)=-\frac{p'(r)}{\rho(r)+p(r)}.
\label{10}
\end{equation}
Finally, comparing equations (\ref{21}) and (\ref{10}), we obtain

\begin{equation}
\frac{dp(r)}{dr} = - \frac{GM(r)\rho(r)}{r^2}\frac{\left( 1 + \frac{p(r)}{\rho(r)} \right) \left( 1 + \frac{4 \pi r^3 p(r)}{M(r)} \right)}{\left( 1 - \frac{2GM(r)}{r} \right)},
\label{8}
\end{equation}  
with the equation (\ref{1}) conveniently rewritten as
\begin{equation}
d M(r)=4 \pi r^{2} \rho(r)dr.
\label{9}
\end{equation}
Equations (\ref{8}) and (\ref{9}) represent the reduction of the Einstein equations into the interior of a symmetric, static and spherical relativistic star. They are the equations of hydrostatic equilibrium for GR, known as TOV \cite{tov1,tov2}.

\subsection{GENERALIZED TOV EQUATIONS}
We now present a generalized version of the TOV based on \cite{Velten2016}. This generalization is given by a free parameterization of the terms of the relativistic stellar hydrostatic equilibrium equations. This parameterization makes it possible to analyze separately the contributions and physical relevance in a clear and objective way of each term present in the TOV, which are originated from relativistic corrections.

\subsection{Parameterized TOV}
\label{ptov}

The parameterization of the TOV (PTOV) equations is characterized by the implementation of five 
new free parameters, namely: $ \alpha $, $ \beta $, $ \gamma $, $ \chi $, and $ \sigma $, where each parameter is related to a particular term that contributes to the TOV equations. 
This procedure incorporate all modified gravity models, like $f(R)$, newtonian and neo-newtonian theories, on a hydrostatic equilibrium equation of a star within the same original TOV structure \cite{Velten2016}.

The parameterization performed (PTOV) is give by,
{\fontsize{9.0}{12}
\begin{equation}
\frac{dp(r)}{dr} = - \frac{G(1+\alpha)\mathcal{M}(r)\rho(r)}{r^2} \frac{\left( 1 + \frac{\beta p(r)}{\rho(r)} \right) \left(1 + \frac{\chi 4 \pi r^3 p(r)}{\mathcal{M}(r)} \right)}{\left( 1 - \frac{\gamma 2G\mathcal{M}(r)}{r} \right)},
\label{5}
\end{equation}
}
where, the mass function $M(r)$ is generalized to an effective mass, written as:
\begin{equation}
\frac{d \mathcal{M}(r)}{dr} = 4\pi r^2 (\rho (r)+\sigma p(r)),
\label{6}
\end{equation}
which is now quite different from the usual mass definition computed in expression (\ref{9}). 
We can observe that for the parameter $\alpha \neq 0$, the definition (\ref{6}) is notoriously different 
from the expression for the common mass calculated according to (\ref{1}). However, the gravitational mass is still obtained from the definition (\ref{4}). Here we briefly discuss the physical interpretation of each of the parameters implemented in the above equations, according to  ref.\cite{Velten2016}:

\begin{itemize}
\item The parameter $ \alpha $ measures the degree of coupling of matter in the stellar object, {\it i.e.}, it is related to possible effects originated from the gravitational coupling $G_{eff} = G(1+\alpha)$. In particular, in GR $ \alpha = 0 $, and in modified gravity theory such as $ f(R)$, $\alpha=1/3$ \cite{Brax2008}.
\end{itemize}

\begin{itemize}
\item $\beta $ measures possible contributions arising from the inertial pressure. See that $ \beta $ in equation (\ref{5}) is located in the contribution of the term $(\rho + p)$ which was originated from the conservation law $\nabla_{\mu}T^{\mu\nu} = 0$ as can be seen in the expression (\ref{10}). This term plays the role of an inert mass density as can be seen in \cite{Schwab2008}. In GR $\beta=1$.
\end{itemize}

\begin{itemize}
\item $\chi$ quantifies the effect of the pressure on self-gravity. This effect appears exclusively in GR and is completely recovered by $\chi=1$.
\end{itemize}

\begin{itemize}
\item The $\gamma$ parameter is related to an intrinsic curvature present in the relativistic context. The term carries a correction originated within the scope of GR according to the proper time definition of a sphere. $\gamma = 0$ corresponds to Newtonian physics, whereas in GR $\gamma = 1$.
\end{itemize}

\begin{itemize}
\item Finally, $\sigma$ modifies the way the $M(r)$ function is computed in (\ref{8}). This parameter is responsible for measuring the effect of the pressure contribution to the gravitational mass of stellar dense objects. In GR $\sigma=0$.
\end{itemize}

Some specific parameter values lead to the recovery of some particular cases, such as: Newtonian hydrostatic, neo-Newtonian hydrodynamic and modified gravity f(R). For more details see \cite{Velten2016}. 

\subsection{GTOV}

We introduce here an alternative version to the previously described PTOV proposal. We generalize the PTOV equations (GTOV) keeping the free parameterization implemented in TOV, and also implementing a second order correction in the continuity equation for the mass. In strong gravitational regimes such as the stellar interior, we assume that equation (\ref{1}) is formally the same as that obtained from a non-relativistic treatment. Thus, the new equation for the mass acquires the following form,

\begin{equation}
\frac{d \mathcal{\tilde{M}}(r)}{dr} = 4\pi r^2 (\rho (r)+\sigma p(r)) + \tilde{\Gamma}(\rho(r))\mathcal{\tilde{M}}(r),
\label{11}
\end{equation}
where $\tilde{\Gamma}(\rho(r)) = \rho^{\frac{1}{2}}(r)\Gamma$ 
varies with the density $\rho$ and depends on the free parameter
$\Gamma$. Note that $\tilde{\Gamma}(\rho(r))$ has $fm^{-1}$ units 
and $\Gamma$ has dimension of $[fm/MeV]^{\frac{1}{2}}$. For $\Gamma
= 0$ we recover Eq. (\ref{6}), i.e., PTOV. In addition, equation (\ref{5}) becomes

{\fontsize{9.4}{12}
\begin{equation}
\frac{dp(r)}{dr} = - \frac{G(1+\alpha)\mathcal{\tilde{M}}(r)\rho(r)}{r^2} \frac{\left( 1 + \frac{\beta p(r)}{\rho(r)} \right) \left(1 + \frac{\chi 4 \pi r^3 p(r)}{\mathcal{\tilde{M}}(r)} \right)}{\left( 1 - \frac{2G\mathcal{\tilde{M}}(r)}{r} \right)}.
\label{12}
\end{equation}
}
As for equation (\ref{11}), the strategy adopted here does not change the way the stellar mass is computed. We want to analyze the possible effects caused in the mass-radius diagram by means of the small modifications inserted in the TOV. Based on this, we must take into account the complications arising from the integration of the $M(r)$ function resulting from these modifications. Regardless of the functional form of the effective mass $\mathcal{\tilde{M}}$, the stellar mass is computed as (\ref{4}), {\it i.e.}, $M = \int_{0}^{R} 4 \pi r^2 \rho(r)dr$. Physically, for $\rho = 0$, (\ref{11}) still satisfies the condition that $\frac{d\mathcal{\tilde{M}}(r)}{dr} = 0$ on the stellar exterior. Such a condition is used in the integration of TOV. This makes it possible for an observer at infinity to make measurements of the actual gravitational mass in the way it is traditionally performed. Note that modified stellar mass equations similar to (\ref{11}) were already used in compact stars studies like \citep{sumanta}. We next consider that the TOV equations assume the new form given by (\ref{11}) and (\ref{12}).

\section{Equation of state \label{sec2}}

In this section we present the equations of state (EoS) used in the present work to 
describe hadronic matter (neutron stars interior) and quark matter (strange star interior).The input to solve the TOV equations comes from the equations of state. 

The hadronic EoS is derived from the Quantum Hadrodynamical (QHD) with nonlinear terms
\cite{livrowalecka,walecka,bogutabodmer}, which is based on a 
relativistic mean field theory and describes the baryon interaction through the exchange of scalar and vector mesons. The first version of the model \cite{Walecka-74} had just the scalar $\sigma$ and vector $\omega$ mesons, which are enough to describe the nuclear saturation. Thereafter the isovector meson $\rho$ was included on the model, which makes it possible 
to describe asymmetric nuclear matter. Finally, to better reproduce the compressibility, effective mass and symmetry energy values, non linear terms on the scalar and vector fields were introduced in the Lagrangian model \cite{bogutabodmer}.

We next use an extended version of the relativistic QHD \cite{Serot}, whose Lagrangian density reads:

\begin{widetext}
\begin{eqnarray}
\mathcal{L}_{QHD} = \sum_B \bar{\psi}_B[\gamma^\mu(i\partial_\mu  - g_{B\omega}\omega_\mu  - g_{B\rho} \frac{1}{2}\vec{\tau} \cdot \vec{\rho}_\mu)
- (m_B - g_{B\sigma}\sigma)]\psi_B  -U(\sigma) +   \nonumber   \\
  + \frac{1}{2}(\partial_\mu \sigma \partial^\mu \sigma - m_s^2\sigma^2) + \frac{1}{4}\Omega^{\mu \nu}\Omega_{\mu \nu} + \frac{1}{2} m_v^2 \omega_\mu \omega^\mu 
+ \frac{1}{2} m_\rho^2 \vec{\rho}_\mu \cdot \vec{\rho}^{ \; \mu} - \frac{1}{4}\bf{P}^{\mu \nu} \cdot \bf{P}_{\mu \nu}  , \label{s1} 
\end{eqnarray}
\end{widetext}
where the sum in $B$ stands for all the baryon octet, $\psi_B$ are the baryonic Dirac fields, and $\sigma$, $\omega_\mu$ and $\vec{\rho}_\mu$ are the mesonic fields. The $g's$ are the Yukawa coupling constants that simulate the strong interaction, $m_B$ is the mass of the baryon $B$, $m_s$, $m_v$ and $m_\rho$ are the masses of the $\sigma$, $\omega$ and $\rho$ mesons respectively. The antisymmetric mesonic field strength tensors are given by their usual expressions as presented in~\cite{gledenning}. The $U(\sigma)$ is the self-interaction term introduced in ref.~\cite{bogutabodmer} to reproduce some of the saturation properties of the nuclear matter and is given by:
 
\begin{equation}
U(\sigma) = \frac{1}{3!}\kappa \sigma^3 + \frac{1}{4!}\lambda \sigma^{4} 
\label{s2} .
\end{equation}

Finally, $\vec{\tau}$ are the Pauli matrices. In order to describe a neutral, chemically stable hypernuclear matter, we add leptons as free Fermi gases:
 
\begin{equation}
\mathcal{L}_{lep} = \sum_l \bar{\psi}_l [i\gamma^\mu\partial_\mu -m_l]\psi_l , 
\label{s3}
\end{equation}
where the sum runs over the two lightest leptons ($e$ and $\mu$).

As discussed in the Introduction, there are many possible parameterizations of the QHD model. We choose two largely used parameterizations to proceed with our studies, namely GM1 and IU-FSU. The GM1 parameterization was proposed by Glendenning \cite{gledenning} specifically to describe stellar matter and presents quite a stiff EOS and the IU-FSU was proposed by Piekarewicz and co-authors \cite{iufsu} and satisfies both nuclear matter \cite{Dutra2014} and astrophysical constraints \cite{Dutra2016} when hyperons are not included. 

There are many approaches to fix the hyperon-meson couplings based either on the adjustments of pure phenomenological potentials \cite{james1,james2} or on group symmetries \cite{lll}. In the present paper, we assume that \cite{Glendenning}:
\begin{equation}
\frac{g_{Y\sigma}}{g_{N\sigma}} = 0.7, \quad \frac{g_{Y\omega}}{g_{N\omega}} = 0.783, \quad \frac{g_{Y\rho}}{g_{N\rho}} = 0.783, 
\label{s16}
\end{equation}
in such a way that $U_\Lambda = -28$ MeV for the GM1 parameterization. It is worth noting that within this parameterization the $\rho$ meson always couples to the isospin projection $I_3$. Nevertheless, the value of the $\rho$-hyperon coupling constants remains arbitrary ~\cite{gledenning}. To avoid discussing this point, which is certainly a matter of huge uncertainties, we use the same values for the IU-FSU hyperon-meson couplings. It is however, important to remember that whichever choice is made, the EOS with hyperons is always softer than the EOS with nucleons only. 

To solve the equations of motion, we use the mean field approximation (MFA), where the meson fields are replaced by their expectation values, i.e:  $\sigma$ $\to$ $\left < \sigma \right >$ = $\sigma_0$,   $\omega^\mu$ $\to$ $\delta_{0 \mu}\left <\omega^\mu  \right >$ = $\omega_{0}$  and   $\rho^\mu$ $\to$ $\delta_{0 \mu}\left <\rho^\mu  \right >$ = $\rho_{0}$. The MFA gives us the following eigenvalue for the baryon energy \cite{gledenning}:

\begin{equation}
E_B = \sqrt{k^2 + M^{*2}_B} + g_{B\omega}\omega_0 + g_{B\rho} \frac{\tau_3}{2}  \rho_0 ,  \label{s4}
\end{equation}
where $M_B^*$ is the baryon effective mass: $M^{*}_B$ $ \dot{=} $ $  m_B - g_{B\sigma}\sigma_0$. 

For the leptons, the energy eigenvalues are those of the free Fermi gas:
\begin{equation}
\quad E_l = \sqrt{k^2 + m^{2}_l} , 
\label{s5}
\end{equation}
and the meson fields become:

\begin{equation}
\omega_0  =\sum_B \frac{g_{B\omega}}{m_v^2} n_B, 
\label{s6}
\end{equation}

\begin{equation}
\rho_0  = \sum_B \frac{g_{B\rho}}{m_\rho^2} \frac{\tau_{3}}{2} n_B,  
\label{s8}
\end{equation}

\begin{equation}
\sigma_0 =  \sum_B \frac{g_{B\sigma}}{m_s^2}  n_{SB} - \frac{1}{2}\frac{\kappa}{m_s^2}\sigma_0^2 -\frac{1}{6}\frac{\lambda}{m_s^2} \sigma_0^3, 
\label{s9}
\end{equation}
where $n_{SB}$ is the scalar density  and $n_{B}$ is the number  density of the baryon $B$:

\begin{align}
\quad n_{SB} &=  \int_0^{k_{FB}} \frac{M^*}{\sqrt{k^2 + M^{*2}}} \frac{k^2}{\pi^2} dk , 
\quad  n_B = \frac{k_{FB}^3}{3\pi^2}, \nonumber \\ 
\quad n &= \sum_B n_B. \label{s10}
\end{align}

To describe the properties of the hypernuclear matter, we calculate the EoS from statistical mechanics~\cite{greiner}. The baryons and leptons, being fermions, obey the Fermi-Dirac distribution. In order to compare our results with experimental and observational constraints, we next study nuclear and stellar systems at zero temperature. In this case the Fermi-Dirac distribution becomes the Heaviside step function. The  energy densities of  baryons, leptons and  mesons (which are bosons) read:

\begin{equation}
\epsilon_B =  \frac{1}{\pi^2} \sum_B \int_0^{k_F} \sqrt{k^2 + M^{*2}_B} k^2 dk, 
\label{s11}
\end{equation}

\begin{equation}
\epsilon_l = \frac{1}{\pi^2} \sum_l \int_0^{k_F} \sqrt{k^2 + m^{2}_l} k^2 dk, 
\label{s12}
\end{equation}

\begin{equation}
\epsilon_m = \frac{1}{2}\bigg ( m_s^2\sigma_0^2 + m_v^2\omega_0^2 + m_\rho^2\rho_0^2 \bigg ) + U(\sigma), \label{s13}
\end{equation}
where $k_F$ is the Fermi momentum, and we have already used the fact that the fermions have degeneracy  equal to 2. The total energy density is the sum of the partial ones:

\begin{equation}
\epsilon = \epsilon_B + \epsilon_l + \epsilon_m,
\label{s14}
\end{equation}
and the pressure is calculated via thermodynamic relations:

\begin{equation}
P = \sum_f \mu_f n_f - \epsilon , 
\label{s15}
\end{equation}
where the sum runs over all the fermions ($f = B,l$) and $\mu$ is the chemical potential, which
corresponds exactly to the energy eigenvalue at $T=0$.

To investigate the quark stars we use a simple relativistic model to describe quark matter, the MIT bag model \cite{mitbag}. The MIT bag model confines the quarks in a volume space delimited by a certain pressure. Inside the bag, a constant positive potential energy per unit volume, namely {\it bag constant (B)}, is necessary so that the bag can be created and kept in the vacuum. Inside this volume, the moving quarks have a kinetic energy and no colour currents survive in the surface. Hence, we assume the quarks in the interior of the bag as a Fermi gas whose energy at the border of the bag is negligible when compared with the energies inside it. The MIT bag model energy density reads
\begin{equation}
\varepsilon = B+\sum_{i} \frac{\gamma_i}{2 \pi^{2}}\int^{k_F}_{0}k^2 (m_i^2 +k^2)^{1/2}dk,
\label{mitenergdens}
\end{equation}
and the pressure is,

\begin{equation}
p = - B +\frac{1}{3}\sum_{i} \frac{\gamma_i}{2 \pi^{2}}\int^{k_F}_{0} 
\frac{k^4}{(m_i^2 +k^2)^{1/2}}dk, 
\label{mitpress}
\end{equation}
where $i=u,d,s$, $\gamma_i$ refers to the degeneracy of the system and accounts for the number of
coulors (3) and the spin (2) and $m_i$ is the quark mass, which we take as $5$ MeV for the $u$ and $d$ quarks and $150$ MeV for the $s$ quark.

Now we can model the hadronic star using the GM1 and IU-FSU parameterizations, solve equations (\ref{s14})
and (\ref{s15}) and use them as input to the TOV equations to obtain as output the mass-radius properties of the star. The same is done with the quark matter to describe quark (or strange) stars,i.e., we solve equations (\ref{mitenergdens}) and (\ref{mitpress}) and use them to obtain the mass-radius properties of a family of quarks stars. In this case, the Bodmer-Witten conjecture \cite{conjecture} is satisfied, as pointed out in \cite{james}.

\section{Results and discussion \label{sec3}}

Along side with PTOV, we now test the contribution of the extra term implemented in equation (\ref{11}) in the analysis of the stellar hydrostatic equilibrium. First, it is necessary to have in hand information about the matter contained within the star. This information is obtained from the equations of state (EoS) presented in the last section. Using a fixed configuration for $\alpha$, $\beta$, $\chi$, $\sigma$ and $\Gamma$, we explore the effect of the new term added to (\ref{11}) on the mass-radius diagram. From now on, the analyses of the parameters refer to PTOV and GTOV only, but the values associated with TOV are also included for the sake of comparison.

\begin{figure*}[t]
\centering
\begin{tabular}{ll}
\includegraphics[width=6.cm,angle=270]{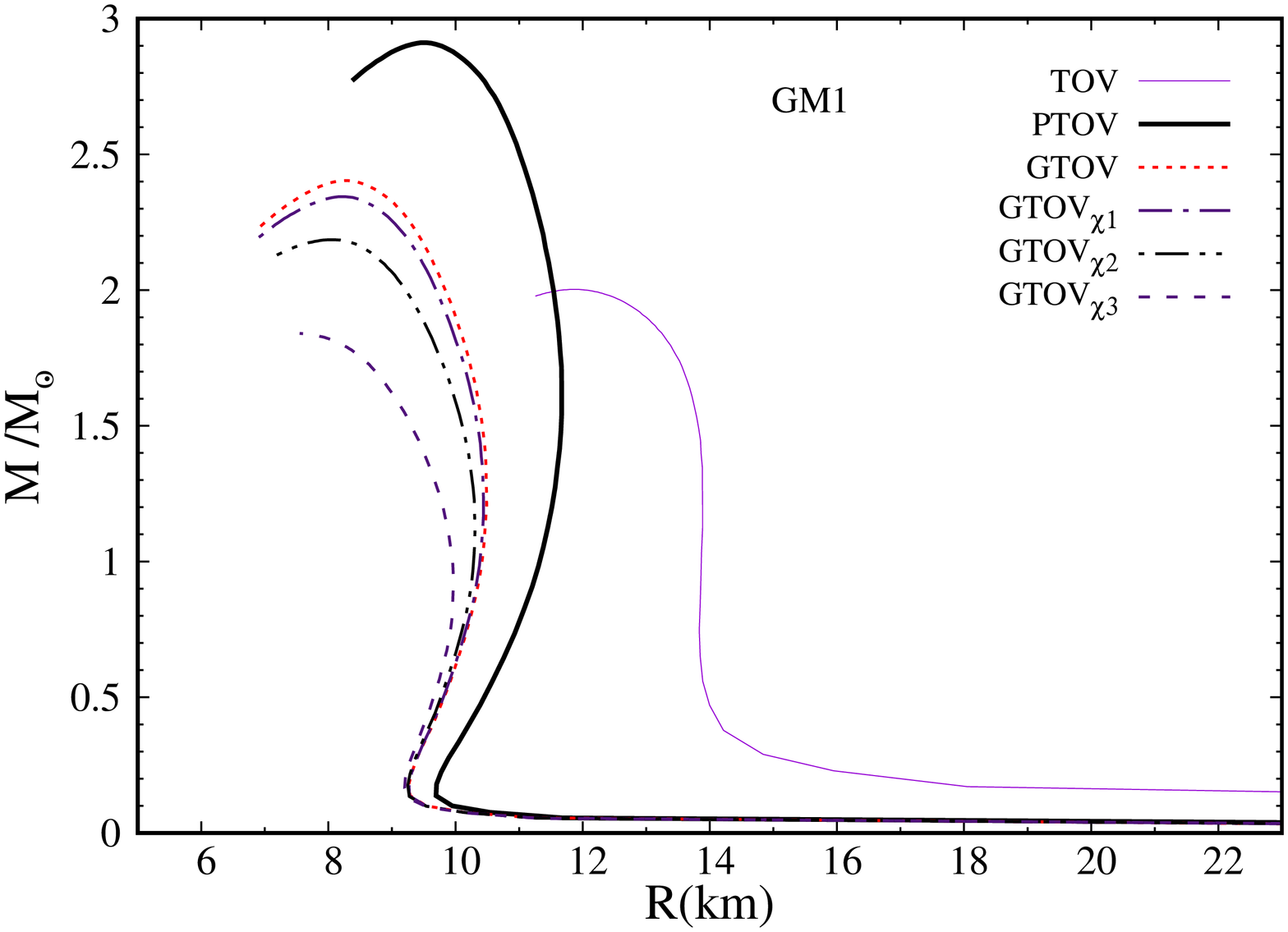} 
\includegraphics[width=6.cm,angle=270]{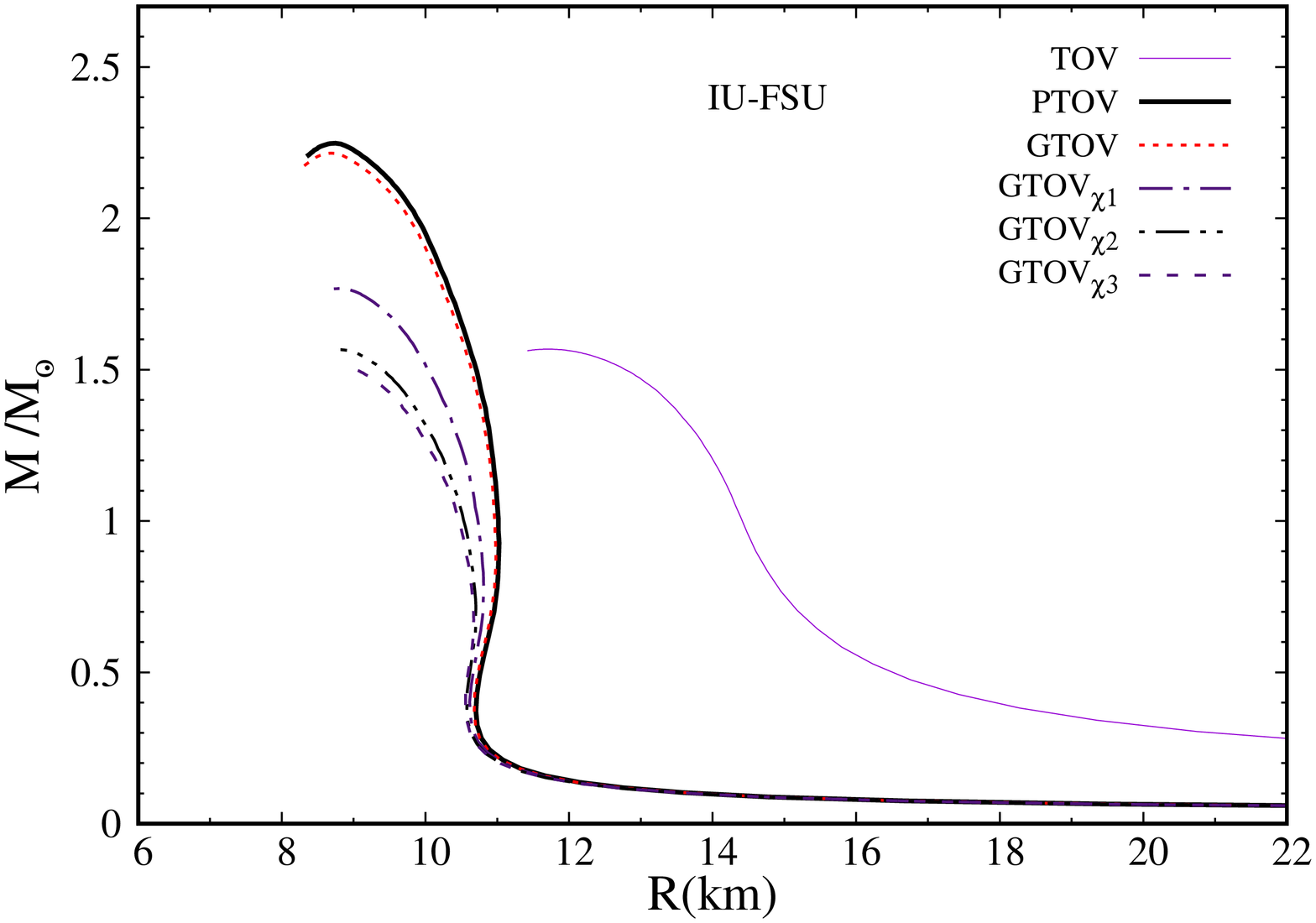}
\end{tabular}
\caption{Mass-radius relation for a family of hadronic stars described with the GM1 EoS (left) and IU-FSU EoS (right). We analyze the effects caused by varying $\chi$ while keeping the other parameters fixed with the values chosen for PTOV.}
\label{fig1}
\end{figure*}

\begin{figure*}[t]
\centering
\begin{tabular}{ll}
\includegraphics[width=6.cm,angle=270]{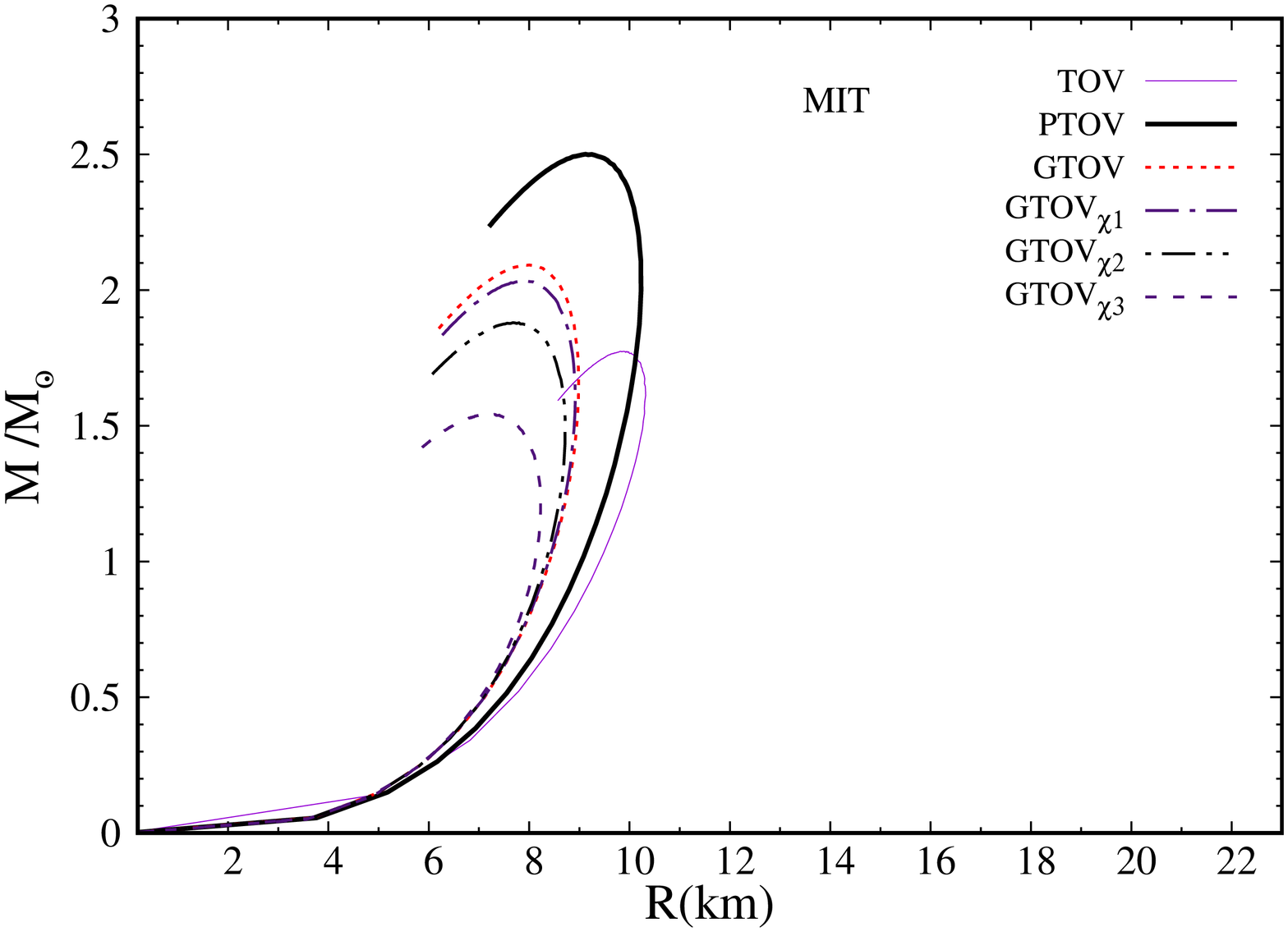} 
\end{tabular}
\caption{Mass-radius relation for a family of quark stars described with the MIT EoS.  We analyze the effects caused by varying $\chi$ while keeping the other parameters fixed with the values chosen for PTOV.}
\label{fig2}
\end{figure*}

\begin{figure*}[!]
\begin{tabular}{ll}
\includegraphics[width=6.cm,angle=270]{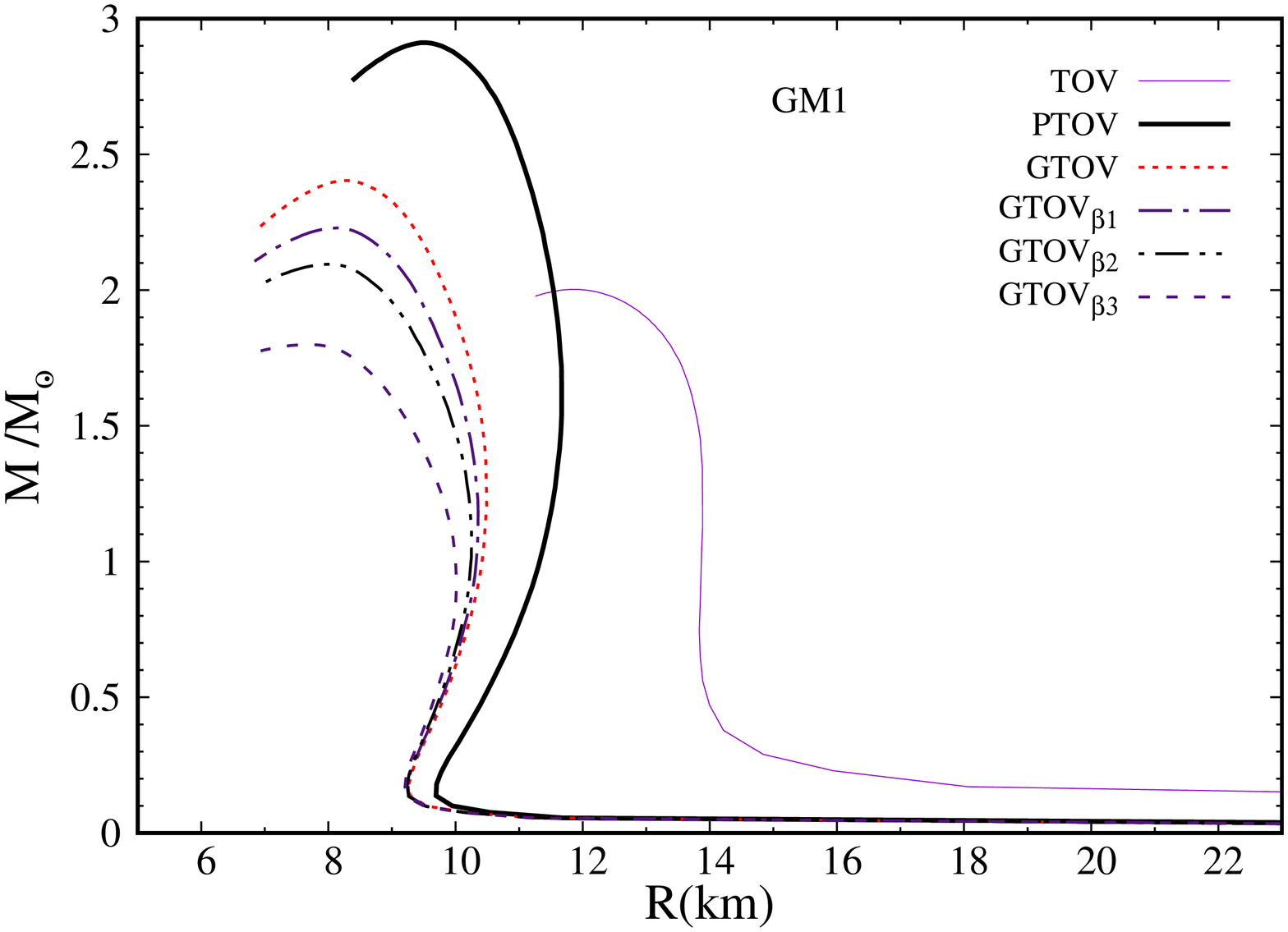} 
\includegraphics[width=6.cm,angle=270]{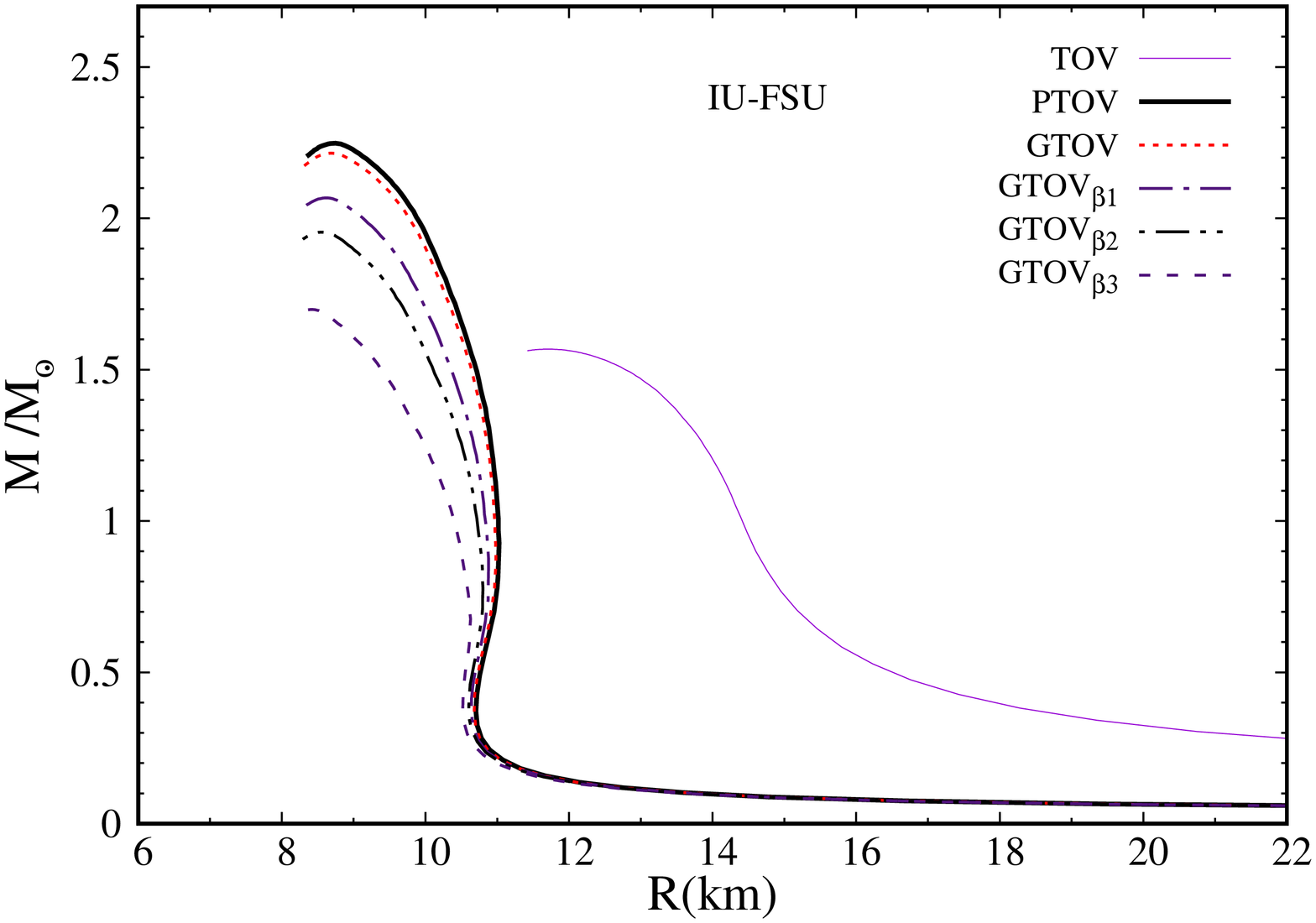}
\end{tabular}
\caption{Mass-radius relation for a family of hadronic stars described with the GM1 EoS (left) and IU-FSU EoS (right). We analyze the effects caused by varying $\beta$ while keeping the other parameters fixed with the values chosen for PTOV.}
\label{fig3}
\end{figure*}

Based on the results obtained in figures (\ref{fig1}), (\ref{fig2}) and (\ref{fig3}), we analyze the effects of the implementation of the correction term in the mass equation together with the modifications
due to the parameterization of the TOV equations.  The effects due to theories of modified gravity are in this context related by the parameter $\alpha$, inserted in the redefinition of the gravitational
coupling $ G_{eff} = G(1+\alpha)$. As $\alpha$ grows, we get equilibrium configurations with the smallest radii and maximum masses. The authors from \cite{Velten2016} tested $\alpha=0$, 1/3 and -1/3. We have tested a large range of values, from -2.0 to 2.0, and the best fit we found was obtained for $\alpha=1.3$ which we maintain fixed in this work. For the TOV case, $\alpha=0$.

To check the effects caused by $\chi$, we vary it within a restricted range, in between -2.0 and 2.0 and the results obtained for hadronic and quark stars are displayed in Tables \ref{tab1} and \ref{tab2} and Figs. \ref{fig1} and  \ref{fig2}. The effects are visible both in the maximum stellar masses and the radii of the star families. While the values used for PTOV dramatically increase the maximum mass, the $\Gamma$ parameter can control the mass while decreasing even further the radii. For greater values of $\chi$ we obtain smaller maximum mass configurations and a small effect on the corresponding radius. This is in line with the results discussed in \cite{Schwab2008}.

\begin{figure*}[t]
\centering
\begin{tabular}{ll}
\includegraphics[width=6.cm,angle=270]{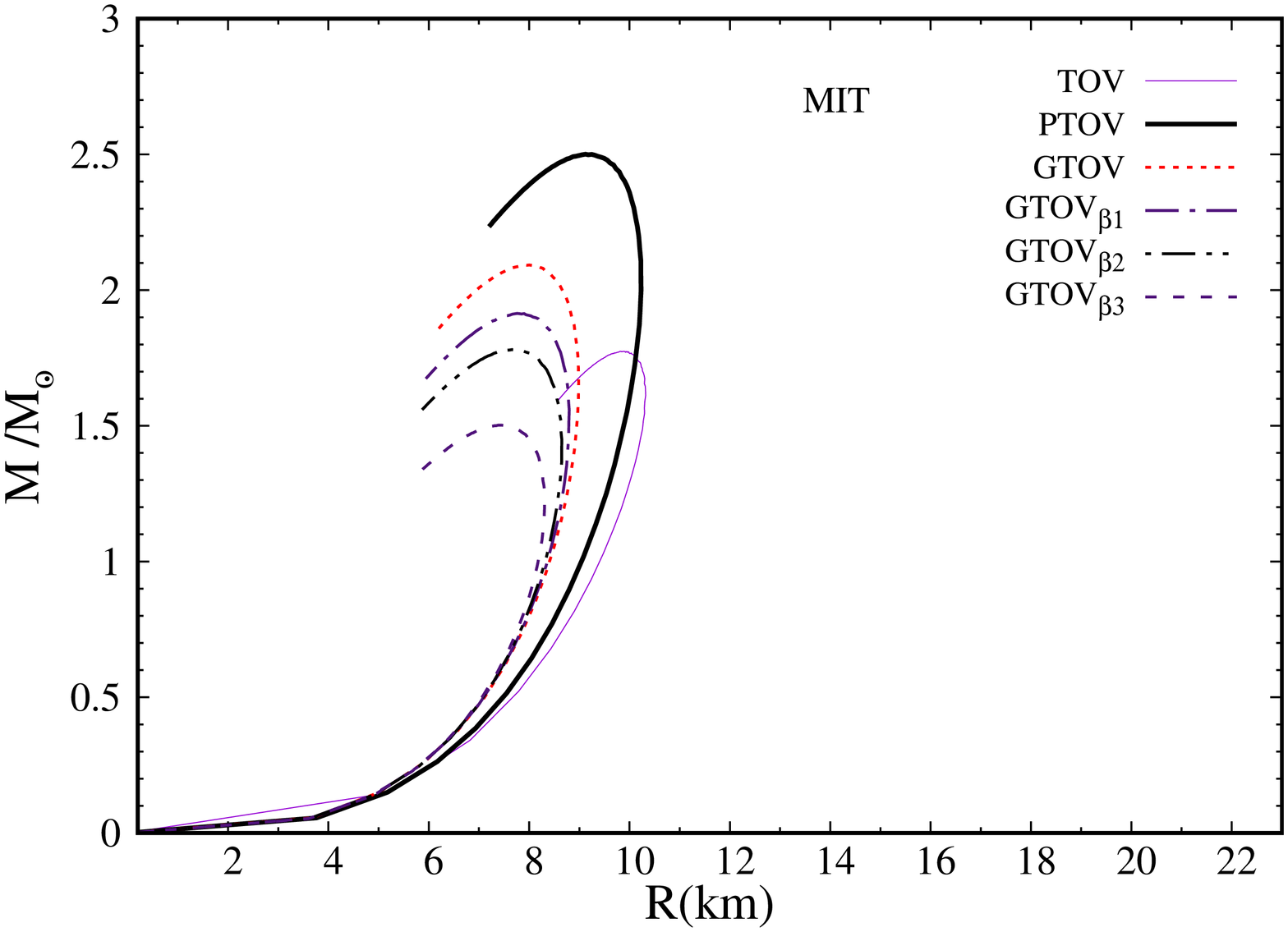} 
\end{tabular}
\caption{Mass-radius relation for a family of quark stars described with the MIT EoS. We analyze the effects caused by varying $\beta$ while keeping the other parameters fixed with the values chosen for PTOV.}
\label{fig4}
\end{figure*}

We next investigate the effects of varying the  $\beta$ parameter in the range -3.0 to 2.0. In
Figs. \ref{fig3} and \ref{fig4} and Tables \ref{tab3} and \ref{tab4}, we display the mass-radius relation for a family of hadronic and quark stars. As $\beta$ grows, there is an increase in the inertia effects
of the pressure resulting in a reduction of the maximum mass of the star. Once again, as masses increase, the corresponding radii decrease. 

The authors from \cite{Velten2016} tested two values for the  $\sigma$ parameter, $\sigma =$ 0 and 3. We have tested a larger range of fraction values, i.e., $\sigma= $ 1/15, 1/8,...,8/3 in between 0 to
3.0, and concluded that the best fit within this range is $\sigma=1/6$.

\begin{figure*}[!]
\begin{tabular}{ll}
\includegraphics[width=6.cm,angle=270]{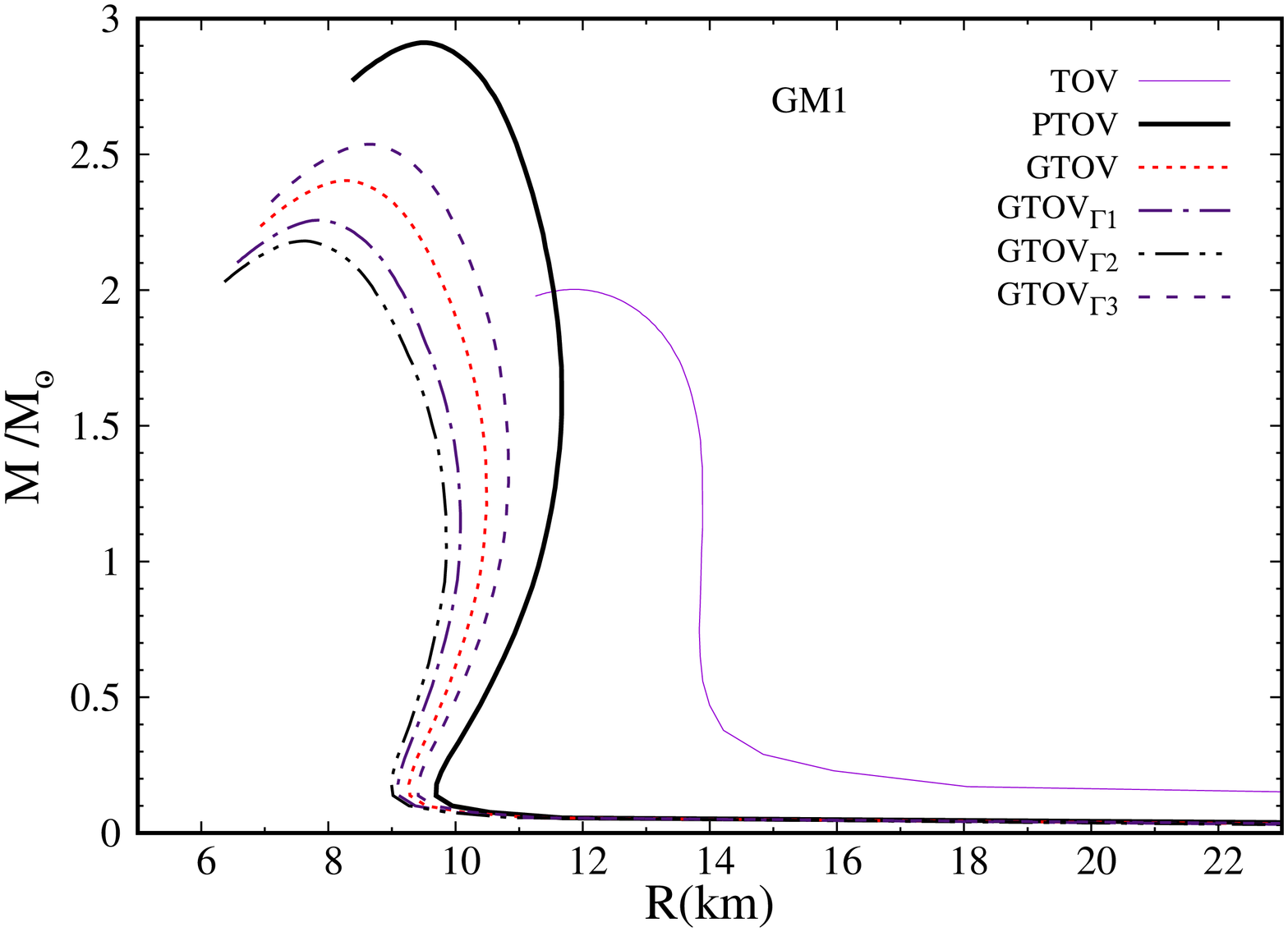} 
\includegraphics[width=6.cm,angle=270]{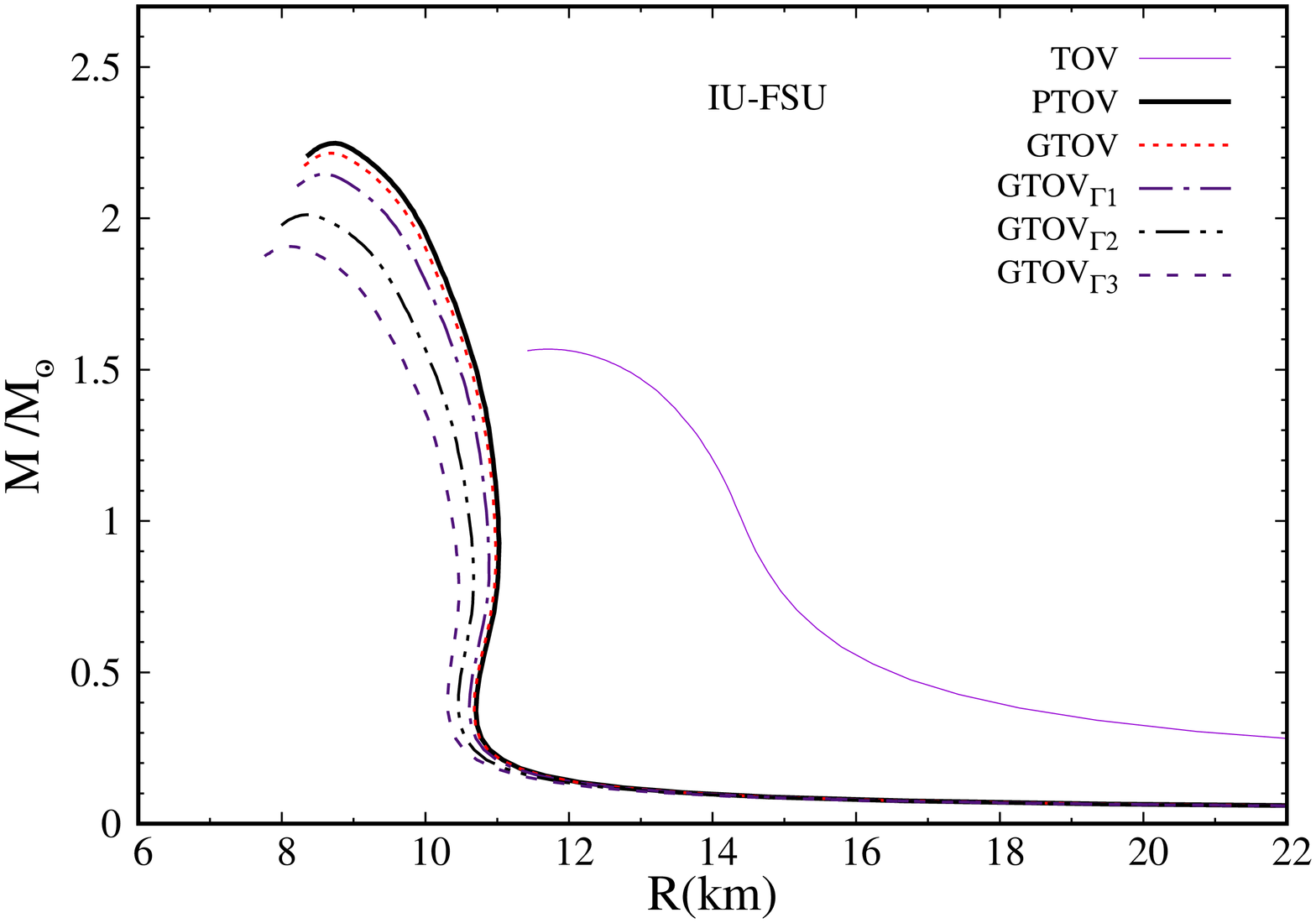}
\end{tabular}
\caption{Mass-radius relation for a family of hadronic stars described with the GM1 EoS (left) and IU-FSU EoS (right). We analyze the effects caused by varying $\Gamma$ while keeping the other parameters fixed with the values chosen for PTOV.}
\label{fig5}
\end{figure*}

\begin{figure*}[t]
\centering
\begin{tabular}{ll}
\includegraphics[width=6.cm,angle=270]{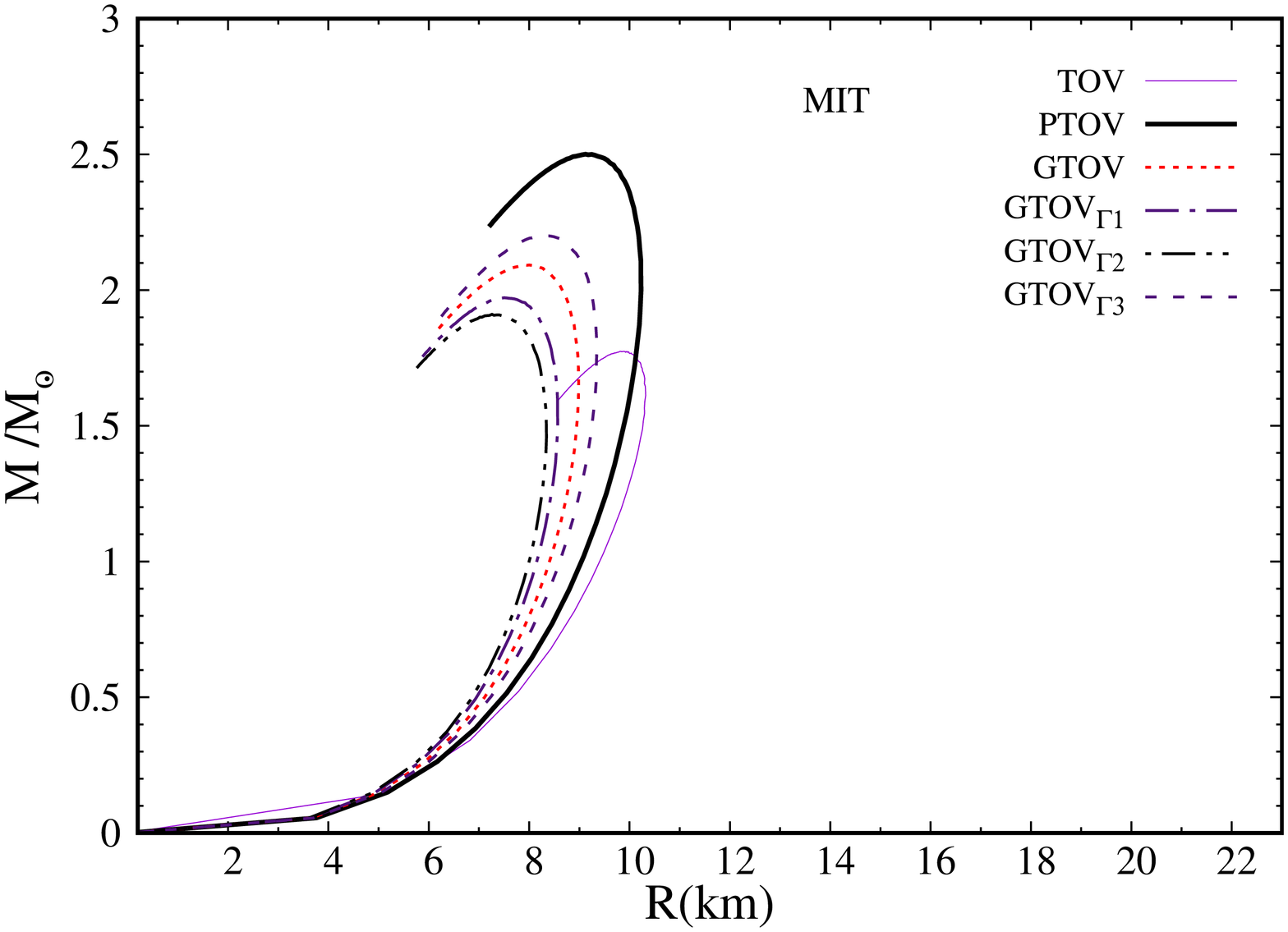} 
\end{tabular}
\caption{Mass-radius relation for a family of quark stars described with the MIT EoS.  We analyze the effects caused by varying $\Gamma$ while keeping the other parameters fixed with the values chosen for PTOV.}
\label{fig6}
\end{figure*}

Finally, we vary $\Gamma$ and analyze the contribution caused due $\tilde{\Gamma}(\rho(r))\mathcal{\tilde{M}}(r)$, ranging from 0 to 4.0. In Figs. \ref{fig5} and \ref{fig6}
the mass-radius relation for a family of hadronic stars and quark stars are shown. The corresponding values of maximum mass and radius, as well as the radius of the canonical 1.4$M_\odot$ neutron star and
quarks stars are written in Tables \ref{tab5} and \ref{tab6}. We note that both, the maximum mass and its radius get smaller with the increase of $\Gamma$. Note that this parameter has a clear effect on the radii of the neutron stars, the star gets a smaller (more compact) radius when we increase the strength of the term. The correction term we introduce in this work makes the gravitational effect stronger and hence, this is an expected result. 

In general, canonical stars, constituted either by hadronic matter or by quark matter present a radius larger than maximum mass stars, as can be easily seen in all the presented figures. Hence, to reconcile a
large maximum mass with a not too large canonical star radius, all the three major ingredients entering the TOV equations (degeneracy pressure, special and general relativity) were revisited and the inclusion of the new term ($\Gamma$) introduced in the present work is essential. Besides the constraints discussed in the Introduction, the merging of two neutron stars belonging to a binary system that resulted in a gravitational wave GW170817, provided also a constraint on the upper limit of neutron star masses of the
order of 2.2 to 2.3 $M_\odot$ \cite{Margalit,Shibata,Rezzolla}. If this constraint is confirmed, many of the the PTOV results are excluded \cite{Velten2016}, but it could be used to establish a well delimited range of values for the $\Gamma$ parameter in GTOV, as can be seen from the values displayed in all Tables.

We now turn our attention to the compactness of the stars ($C_{M_{\text{max}}}$ for the case $M=M_{\text{max}}$ and $C_{1.4M_\odot}$ for $M=1.4~M_\odot$), defined as the ratio between
the masses and corresponding radii of the compact stars. The compactness of a recently measured isolated neutron star \cite{comp} is equal to $0.105 \pm 0.002$. If we analyze all the results displayed
in all Tables, we see that hadronic massive stars tend to be more compact than their canonical counterparts and canonical stars constituted by quarks are more compact than their hadronic counterparts. Had we decided to use compactness of this specific isolated star as an astrophysical constraint, only the 1.4 $M_\odot$ star obtained as a solution from the TOV equations would be acceptable. Nevertheless, more observational results have to be announced before compactness can be used as a viable constraint.    

As a summary, we can say that we have found appropriate parameterizations of the TOV equations, based on realistic physical assumptions, named GTOV, with which all astrophysical constraints proposed so far can be satisfied. As compared with the results presented in \cite{Velten2016}, while maintaining maximum masses high, the radii can decrease in such a way that all the values proposed in \cite{Hebeler2013,Steiner2013,Ozel2016,cskyrme,Ligo2} can be attained. In our prescription, a new dimensionful term that controls the effective mass of the compact object was included and it contributes to the adjustment of the compactness of the stars. 


\begin{table*}[t]
\begin{center}
\begin{tabular}{c|c|cccccc}
\hline
\hline
& Model & \ \ TOV \ \ \ & PTOV \ \ \ & GTOV \ \ \ & GTOV$_{\chi1}$ \ \ \ & GTOV$_{\chi2}$ \ \ \ & GTOV$_{\chi3}$ \tabularnewline 

\hline

 & $\beta$   &  1   & -2.1  & -2.1  & -2.1  & -2.1 & -2.1 \tabularnewline

Parameters & \textbf{$\chi$} & \textbf{1} & \textbf{-1.1} & \textbf{-1.1} & \textbf{-1} & \textbf{-0.7} & \textbf{0.2} \tabularnewline

& $\Gamma$ &  0  &  0  &  2.26  &  2.26  &  2.26  &  2.26 \tabularnewline

\hline

& $M_{max}$  &  2.0 $M_\odot$ &  2.91 $M_\odot$  & 2.40 $M_\odot$  &  2.34 $M_\odot$  & 2.18 $M_\odot$ & 1.84 $M_\odot$ \tabularnewline

Hadronic  & $R_{M_{max}}$ &  11.86 km  \ \ &  9.52 km   \ \ & 8.25 km \ \ & 8.22 km  \ \ & 8.05 km  \ \ & 7.54 km \tabularnewline

stars &$R_{1.4}$& 13.86 km  \ \ &  11.62 km \ \ & 10.46 km  \ \ & 10.40 km  \ \ & 10.19 km  \ \ & 9.49 km \tabularnewline

(GM1)& $C_{M_{max}}$  &  0.16  \ \ &  0.30  \ \ & 0.29  \ \ & 0.28 \ \ & 0.27 \ \ & 0.24 \tabularnewline

& $C_{1.4M_\odot}$  & 0.10  \ \ &  0.12  \ \ & 0.13  \ \ & 0.13 \ \ & 0.14 \ \ & 0.15 \tabularnewline

\hline

 & $M_{max}$ &  1.77 $M_\odot$  \ \ &  2.50 $M_\odot$  \ \ & 2.09 $M_\odot$  \ \ &  2.03 $M_\odot$ \ \ & 1.87 $M_\odot$  \ \ & 1.54 $M_\odot$ \tabularnewline

 Quark & $R_{M_{max}}$ &  9.87 km  \ \ &  9.12 km  & 7.90 km  & 7.88 km  &  7.67 km & 7.14 km \tabularnewline

stars &$R_{1.4}$ & 10.15 km  \ \ &  9.76 km  & 8.88 km  & 8.84 km & 8.70 km & 8.07 km \tabularnewline

(MIT) & $C_{M_{max}}$  &  0.17  \ \ &  0.27  \ \ & 0.26  \ \ & 0.25 \ \ & 0.24 \ \ & 0.21 \tabularnewline

& $C_{1.4M_\odot}$  & 0.13  \ \ &  0.14  \ \ & 0.15  \ \ & 0.15 \ \ & 0.16 \ \ & 0.17 \tabularnewline
\hline
\hline
\end{tabular}
\caption{Values of the $\beta$, $\chi$ and $\Gamma$ parameters corresponding to the mass-radius diagram in Fig.\ref{fig1} (left) and Fig.\ref{fig2}.}
\label{tab1} 
\end{center}
\end{table*}



\begin{table*}[t]
\begin{center}
\begin{tabular}{c|c|cccccc}
\hline
\hline
& Model & \ \ TOV \ \ \ & PTOV \ \ \ & GTOV \ \ \ & GTOV$_{\chi1}$ \ \ \ & GTOV$_{\chi2}$ \ \ \ & GTOV$_{\chi3}$ \tabularnewline 

\hline

 & $\beta$   &  1   & -2.1  & -2.1  & -2.1  & -2.1 & -2.1 \tabularnewline

Parameters & \textbf{$\chi$} & \textbf{1} & \textbf{-1.7} & \textbf{-1.7} & \textbf{-1.3} & \textbf{-1.0} & \textbf{-0.9} \tabularnewline

& $\Gamma$ &  0  &  0  &  0.09  &  0.09  &  0.09  &  0.09 \tabularnewline

\hline

& $M_{max}$  &  1.56 $M_\odot$ &  2.24 $M_\odot$  & 2.21 $M_\odot$  &  1.76 $M_\odot$  & 1.56 $M_\odot$ & 1.51 $M_\odot$ \tabularnewline

 Hadronic & $R_{M_{max}}$ &  11.70 km  \ \ &  8.74 km   \ \ & 8.70 km \ \ & 8.80 km  \ \ & 8.81 km  \ \ & 8.84 km \tabularnewline

stars &$R_{1.4}$& 13.35 km  \ \ &  10.80 km \ \ & 10.73 km  \ \ & 10.24 km  \ \ & 9.78 km \ \ & 9.60 km \tabularnewline

 (IU-FSU) & $C_{M_{max}}$  &  0.13  \ \ &  0.26  \ \ & 0.25  \ \ & 0.2 \ \ & 0.18 \ \ & 0.17 \tabularnewline

& $C_{1.4M_\odot}$  &  0.10  \ \ &  0.12  \ \ & 0.13  \ \ & 0.13 \ \ & 0.14 \ \ & 0.15 \tabularnewline

\hline
\hline
\end{tabular}
\caption{Values of the $\beta$, $\chi$ and $\Gamma$ parameters corresponding to the mass-radius diagram in Fig.\ref{fig1} (right).}
\label{tab2} 
\end{center}
\end{table*}

\begin{table*}[t]
\begin{center}
\begin{tabular}{c|c|cccccc}
\hline
\hline
& Model & \ \ TOV \ \ \ & PTOV \ \ \ & GTOV \ \ \ & GTOV$_{\beta1}$ \ \ \ & GTOV$_{\beta2}$ \ \ \ & GTOV$_{\beta3}$  \tabularnewline 

\hline

& \textbf{$\beta$}  &  \textbf{1}   & \textbf{-2.1}  & \textbf{-2.1}  & \textbf{-1.5}  &  \textbf{-1}  & \textbf{0.3} \tabularnewline

Parameters & $\chi$   &  1  & -1.1  &  -1.1  &  -1.1  & -1.1  &  -1.1 \tabularnewline

& $\Gamma$ &  0  &  0  &  2.26  &  2.26  &  2.26  &  2.26 \tabularnewline

\hline

& $M_{max}$  &  2.0 $M_\odot$ &  2.91 $M_\odot$  & 2.40 $M_\odot$  &  2.22 $M_\odot$  & 2.09 $M_\odot$ & 1.79 $M_\odot$ \tabularnewline

 Hadronic & $R_{M_{max}}$ &  11.86 km  \ \ &  9.52 km   \ \ & 8.25 km \ \ & 8.13 km  \ \ & 8.00 km  \ \ & 7.67 km \tabularnewline

stars & $R_{1.4}$& 13.86 km  \ \ &  11.62 km \ \ & 10.46 km  \ \ & 10.26 km  \ \ & 10.08 km  \ \ & 9.50 km \tabularnewline

(GM1) & $C_{M_{max}}$  &  0.16  \ \ &  0.30  \ \ & 0.29  \ \ & 0.27 \ \ & 0.26 \ \ & 0.23 \tabularnewline

& $C_{1.4M_\odot}$  &  0.10 \ \ &  0.12  \ \ & 0.13  \ \ & 0.13 \ \ & 0.14 \ \ & 0.14 \tabularnewline

\hline

& $M_{max}$ &  1.77 $M_\odot$  \ \ &  2.50 $M_\odot$  \ \ & 2.09 $M_\odot$  \ \ &  1.91 $M_\odot$ \ \ & 1.78 $M_\odot$  \ \ & 1.50 $M_\odot$ \tabularnewline

Quark & $R_{M_{max}}$ &  9.87 km  \ \ &  9.12 km  & 7.90 km  & 7.75 km  &  7.65 km & 7.41 km \tabularnewline

 stars & $R_{1.4}$ & 10.15 km  \ \ &  9.76 km  & 8.88 km  & 8.76 km & 8.65 km & 8.15 km \tabularnewline

(MIT) & $C_{M_{max}}$  &  0.17  \ \ &  0.27  \ \ & 0.26  \ \ & 0.24 \ \ & 0.23 \ \ & 0.20 \tabularnewline

& $C_{1.4M_\odot}$  &  0.13  \ \ &  0.14  \ \ & 0.15  \ \ & 0.16 \ \ & 0.16 \ \ & 0.17 \tabularnewline
\hline
\hline
\end{tabular}
\caption{Values of the $\beta$, $\chi$ and $\Gamma$ parameters corresponding to the mass-radius diagram in Fig.\ref{fig3} (left) and Fig.\ref{fig4}.}
\label{tab3}
\end{center}
\end{table*}



\begin{table*}[t]
\begin{center}
\begin{tabular}{c|c|cccccc}
\hline
\hline
& Model & \ \ TOV \ \ \ & PTOV \ \ \ & GTOV \ \ \ & GTOV$_{\beta1}$ \ \ \ & GTOV$_{\beta2}$ \ \ \ & GTOV$_{\beta3}$  \tabularnewline 

\hline

& \textbf{$\beta$}  &  \textbf{1}   & \textbf{-2.1}  & \textbf{-2.1}  & \textbf{-1.5}  &  \textbf{-1}  & \textbf{0.3} \tabularnewline

Parameters & $\chi$   &  1  & -1.7  &  -1.7  &  -1.7  & -1.7  &  -1.7 \tabularnewline

& $\Gamma$ &  0  &  0  &  0.09  &  0.09  &  0.09  &  0.09 \tabularnewline

\hline

& $M_{max}$  &  1.56 $M_\odot$ &  2.24 $M_\odot$  & 2.21 $M_\odot$  &  2.06 $M_\odot$  & 1.95 $M_\odot$ & 1.69 $M_\odot$ \tabularnewline

 Hadronic & $R_{M_{max}}$ &  11.70 km  \ \ &  8.74 km   \ \ & 8.70 km \ \ & 8.60 km  \ \ & 8.57 km  \ \ & 8.40 km \tabularnewline

stars &$R_{1.4}$& 13.35 km  \ \ &  10.80 km \ \ & 10.73 km  \ \ & 10.49 km  \ \ & 10.29 km \ \ & 9.65 km \tabularnewline

 (IU-FSU) & $C_{M_{max}}$  &  0.13  \ \ &  0.26  \ \ & 0.25  \ \ & 0.23 \ \ & 0.22 \ \ & 0.20 \tabularnewline

& $C_{1.4M_\odot}$  &  0.10  \ \ &  0.12  \ \ & 0.13  \ \ & 0.13 \ \ & 0.13 \ \ & 0.14 \tabularnewline

\hline
\hline
\end{tabular}
\caption{Values of the $\beta$, $\chi$ and $\Gamma$ parameters corresponding to the mass-radius diagram in Fig.\ref{fig3} (right).}
\label{tab4}
\end{center}
\end{table*}

\begin{table*}[t]
\begin{center}
\begin{tabular}{c|c|ccccccc}
\hline
\hline
& Model & \ \ TOV \ \ \ & PTOV \ \ \ & GTOV \ \ \ & GTOV$_{\Gamma1}$ \ \ \ & GTOV$_{\Gamma2}$ \ \ \ & GTOV$_{\Gamma3}$ \tabularnewline 

\hline

& $\beta$   &  1   & -2.1  & -2.1  & -2.1  & -2.1  & -2.1 \tabularnewline

Parameters & $\chi$   &  1  & -1.1  &  -1.1  &  -1.1  & -1.1  &  -1.1 \tabularnewline

& \textbf{$\Gamma$}  &  \textbf{0}  &  \textbf{0}  &  \textbf{2.26}  &  \textbf{3.23}  &  \textbf{3.82}  & \textbf{1.52} \tabularnewline

\hline

& $M_{max}$ &  2.0 $M_\odot$ &  2.91 $M_\odot$  & 2.40 $M_\odot$  &  2.25 $M_\odot$  & 2.18 $M_\odot$ & 2.53 $M_\odot$ \tabularnewline

Hadronic & $R_{M_{max}}$ &  11.86 km  \ \ &  9.52 km   \ \ & 8.25 km \ \ & 7.85 km  \ \ & 7.60 km  \ \ & 8.63 km \tabularnewline

 stars & $R_{1.4}$& 13.86 km  \ \ &  11.62 km \ \ & 10.46 km  \ \ & 10.00 km  \ \ & 9.73 km  \ \ & 10.82 km  \tabularnewline

(GM1) & $C_{M_{max}}$  &  0.16  \ \ &  0.30  \ \ & 0.29  \ \ & 0.28 \ \ & 0.28 \ \ & 0.29 \tabularnewline

& $C_{1.4M_\odot}$  &  0.10  \ \ &  0.12  \ \ & 0.13  \ \ & 0.14 \ \ & 0.14 \ \ & 0.13 \tabularnewline

\hline

& $M_{max}$ &  1.77 $M_\odot$  \ \ &  2.50 $M_\odot$  \ \ & 2.09 $M_\odot$  \ \ &  1.97 $M_\odot$ \ \ & 1.91 $M_\odot$  \ \ & 2.20 $M_\odot$  \tabularnewline

Quark & $R_{M_{max}}$ &  9.87 km  \ \ &  9.12 km  & 7.90 km  & 7.50 km  &  7.26 km & 8.22 km \tabularnewline

 stars & $R_{1.4}$ & 10.15 km  \ \ &  9.76 km  & 8.88 km  & 8.54 km & 8.33 km & 9.16 km \tabularnewline

(MIT) & $C_{M_{max}}$  &  0.17  \ \ &  0.27  \ \ & 0.26  \ \ & 0.26 \ \ & 0.26 \ \ & 0.27 \tabularnewline

& $C_{1.4M_\odot}$  &  0.13 \ \ &  0.14 \ \ & 0.15  \ \ & 0.16 \ \ & 0.17 \ \ & 0.15 \tabularnewline

\hline
\hline
\end{tabular}
\caption{Values of the $\beta$, $\chi$ and $\Gamma$ parameters corresponding to the mass-radius diagram in Fig.\ref{fig5} (left) and Fig.\ref{fig6}.}
\label{tab5}
\end{center}
\end{table*}



\begin{table*}[t]
\begin{center}
\begin{tabular}{c|c|ccccccc}
\hline
\hline
& Model & \ \ TOV \ \ \ & PTOV \ \ \ & GTOV \ \ \ & GTOV$_{\Gamma1}$ \ \ \ & GTOV$_{\Gamma2}$ \ \ \ & GTOV$_{\Gamma3}$ \tabularnewline 

\hline

& $\beta$   &  1   & -2.1  & -2.1  & -2.1  & -2.1  & -2.1 \tabularnewline

Parameters & $\chi$   &  1  & -1.7  &  -1.7  &  -1.7  & -1.7  &  -1.7 \tabularnewline

& \textbf{$\Gamma$}  &  \textbf{0}  &  \textbf{0}  &  \textbf{0.09}  &  \textbf{0.3}  &  \textbf{0.8}  & \textbf{1.3} \tabularnewline

\hline

& $M_{max}$  &  1.56 $M_\odot$ &  2.24 $M_\odot$  & 2.21 $M_\odot$  &  2.14 $M_\odot$  & 2.01 $M_\odot$ & 1.90 $M_\odot$ \tabularnewline

 Hadronic & $R_{M_{max}}$ &  11.70 km  \ \ &  8.74 km   \ \ & 8.70 km \ \ & 8.59 km  \ \ & 8.37 km  \ \ & 8.12 km \tabularnewline

stars &$R_{1.4}$& 13.35 km  \ \ &  10.80 km \ \ & 10.73 km  \ \ & 10.60 km  \ \ & 10.26km \ \ & 9.93 km \tabularnewline

 (IU-FSU) & $C_{M_{max}}$  &  0.13  \ \ &  0.26  \ \ & 0.25  \ \ & 0.25 \ \ & 0.24 \ \ & 0.23 \tabularnewline

& $C_{1.4M_\odot}$  &  0.10  \ \ &  0.12  \ \ & 0.13  \ \ & 0.13 \ \ & 0.14 \ \ & 0.14 \tabularnewline

\hline
\hline
\end{tabular}
\caption{Values of the $\beta$, $\chi$ and $\Gamma$ parameters corresponding to the mass-radius diagram in Fig.\ref{fig5} (right).}
\label{tab6}
\end{center}
\end{table*}


\acknowledgments
This work is a part of the project CNPq-INCT-FNA Proc. No. 464898/2014-5.DPM acknowledge partial support from CNPq (Brazil) and CEM has a scholarship paid by Capes (Brazil).

\end{document}